\documentclass[twocolumn,prl,aps,floatfix]{revtex4} 
\usepackage{times}
\usepackage{graphicx}
\usepackage{dcolumn}
\usepackage{bm}

\begin{document}

\preprint{APS/123-QED}

\title{Diamagnetism of YBa$_{2}$Cu$_{3}$O$_{6+x}$ crystals above $T_c$~: evidence for Gaussian
 fluctuations }

\author{I. Kokanovi\'{c}$^{1,2}$, D. J. Hills$^{1}$,
M. L.  Sutherland$^{1}$,  R. Liang$^{3}$ and J. R. Cooper$^1$}

\affiliation{$^1$Cavendish Laboratory, University of Cambridge,
Cambridge CB3 0HE, U.K.\\
$^2$Department of Physics, Faculty of Science, University of
Zagreb, P.O.Box 331, Zagreb, Croatia.\\ $^3$Department of Physics
and Astronomy, University of British Columbia, Vancouver, British
Columbia, Canada V6T 1Z1}
 \altaffiliation{Electronic address: kivan@phy.hr}



\date{\today}

 \begin{abstract}
The magnetization of~three high-quality single crystals of~YBa$_{2}$Cu$_{3}$O$_{6+x}$,
from slightly overdoped to heavily underdoped,
 has been measured using torque magnetometry. Striking effects in the angular dependence
 of the torque for the two underdoped crystals, a few degrees above the superconducting
 transition temperature ($T_c$)
 are described well by the theory of Gaussian superconducting fluctuations using a single adjustable parameter.
 The data at higher temperatures ($T$) are consistent with a strong cut-off in the fluctuations for
 $T\gtrsim1.1T_c$. Numerical estimates suggest that inelastic scattering could be responsible for this
 cut-off.

 \end{abstract}

\maketitle

Cuprate superconductors show much stronger thermodynamic fluctuations than classical
ones
  because of their higher
   transition temperatures ($T_c$),  shorter Ginzburg-Landau (GL) coherence
lengths
   and  quasi-two dimensional  layered structures with
weakly interacting CuO$_2$
   planes \cite{Bulaevskii,Larkin}.  Observations of diamagnetism  \cite{LuLi} and large Nernst coefficients over a broad
temperature ($T$) range  well above
  $T_c$  for several types of cuprate
 \cite{Xu,Wang06} are intriguing \cite{Kivelson}. They are often cited as evidence for
 pre-formed Cooper pairs without the long-range phase coherence
needed for superconductivity. In contrast, in Ref.~\onlinecite{Tallon2011} it is
argued that phase and amplitude fluctuations set in simultaneously.  However  the
fluctuations are still considered  to be strong in that the mean-field transition
temperature $T_c^{MF}$, obtained  by applying entropy and free energy balance
considerations to heat capacity data, is substantially larger than $T_c$ especially
for underdoped cuprates. In standard GL theory the coefficient of the $|\psi|^2$ term
in the free energy, where $\psi$ is the order parameter, changes sign at $T_c^{MF_1}$,
as explained in footnote~\onlinecite{Lonzarich}.   If $|\psi|^4$ and higher order
terms are neglected, $T_c^{MF_1}$ can  be obtained from
 a Gaussian fluctuation (GF) analysis of the magnetic susceptibility and other physical
properties \cite{Bulaevskii}.

 One difficulty in this area is separating the fluctuation (FL) contribution
to a given property from the normal state (N) background. Recently this has been dealt
with for the in-plane electrical conductivity $\sigma_{ab}(T)$ of
YBa$_{2}$Cu$_{3}$O$_{6+x}$ crystals by applying very high magnetic fields ($B$)
\cite{Alloul}. When analyzed using GF theory, $\sigma_{ab}^{FL}(T)$ was found to cut
off even more rapidly above  $T\gtrsim1.1T_c$ than previously thought
\cite{Genova,Vidal}. It was also strongly reduced at high $B$ and the fields
 needed to suppress $\sigma_{ab}^{FL}(T)$  extrapolated
to zero between 120 and 140 K depending on $x$, which tends to support a vortex or
Kosterlitz-Thouless scenario.   Therefore questions such as
  the applicability of GF theory $vs.$ a phase fluctuation or mobile vortex scenario and the extent to which  $T_c$ is
  suppressed below $T_c^{MF_1}$ by strong critical fluctuations, are still
being discussed. They are of general interest because superconducting fluctuations
could limit the maximum $T_c$ that can be obtained in a given class of material
\cite{Tallon2011}, and moreover  \cite{Alloul} the  fluctuation cut-off could be
linked in some way to the pairing mechanism.

Here we report torque magnetometry data measured \cite{exptldetails} from $T_c$ to 300
K  for tiny  YBa$_{2}$Cu$_{3}$O$_{6+x}$~(YBCO) single crystals from overdoped (OD) to
heavily underdoped (UD). These were grown in non-reactive
  BaZrO$_3$ crucibles from high-purity
(5N) starting materials. Evidence for the quality of the UD crystals includes
extremely sharp x-ray peaks \cite{Liang2000}, and substantial mean free paths from
quantum oscillation measurements \cite{Audouard2009}. The OD89 crystal is from another
preparation batch  which had narrow superconducting transitions and a maximum $T_c$ of
93.8~K \cite{Kirby2005}. We analyze the results using GF theory which, unlike some
other approaches, predicts the \emph{magnitude} of the observed effects as well as
their $T$-dependence. We show that it gives excellent single-parameter fits to the
striking angular dependence of the torque, which has previously been attributed to the
presence of a very large magnetic field scale \cite{LuLi}.  We also show that
inelastic scattering is a plausible mechanism for cutting off the fluctuations at
higher $T$ and a possible alternative to strong fluctuations for limiting $T_c$.

 Although measurements of the London penetration depth
\cite{Kamal} below $T_c$ and thermal expansion \cite{Meingast} above and below $T_c$
for optimally doped (OP) YBCO crystals, give evidence for critical fluctuations
described by the 3D-XY model, up to $\pm$ 10 K from $T_c$, we argue later that these
do not alter our overall picture.

 A crystal with magnetization $M$ in an applied magnetic field $B$  attached to a piezoresistive cantilever causes a change
in electrical resistance proportional to the torque density
$\tau\equiv\underline{M}\times\underline{B}$. If $B$ is parallel to the $c$-axis of a
cuprate crystal, then in the
  low field limit
  the contribution to $M$ in the $c$-axis direction from Gaussian fluctuations ($M_c^{FL}$)  is given by \cite{Larkin}:
\begin{equation}
{M_c^{FL}(T)=-\frac{\pi k_BTB}{3 \Phi_0^2}\frac{
\xi_{ab}^{2}(T)}{s\sqrt{1+[2\xi_{ab}(T)/\gamma s)]^2}}} \label{1}
\end{equation}
Here $\gamma = \xi_{ab}(T)/\xi_{c}(T)$ is the anisotropy, defined as the ratio of the
$T$-dependent coherence lengths $\parallel$ and $\bot$ to the layers, i.e.
$\xi_{ab,c}(T) = \xi_{ab,c}(0)/\epsilon^{1/2}$ with $\epsilon=\ln(T/T_c^{MF_1})$
\cite{Larkin,Alloul}. The distance between the CuO$_2$ bi-layers is taken as $s$ =
1.17 nm, and $ \Phi_0$ and $k_B$ are the pair flux quantum and Boltzmann's constant
respectively. For $B\perp c$ the fluctuation magnetization is negligibly small.

As the angle $\theta$ between the applied field and CuO$_2$ planes is altered,
$\tau(\theta)$ will vary as $\tau(\theta)= \frac{1}{2}\chi_D(T)B^2\sin2\theta$, as
long as $M\propto B$. Thus, fits to $\tau(\theta) \propto B^2\sin2\theta$ give
$\chi_D(T)\equiv\chi_c(T)-\chi_{ab}(T)$, which is the susceptibility anisotropy.
Fig.~1 shows torque data for UD57 up to 15~K above the low-field $T_c$ of 57~K. Much
of our data, including the two curves for UD57 in Fig.~1 at higher $T$ follow a
$\sin2\theta$ dependence very closely,  however there are striking deviations at lower
$T$ arising from non-linearity in $M(B)$ that we discuss later.

 \begin{figure}
\includegraphics[width=7.0cm,keepaspectratio=true]{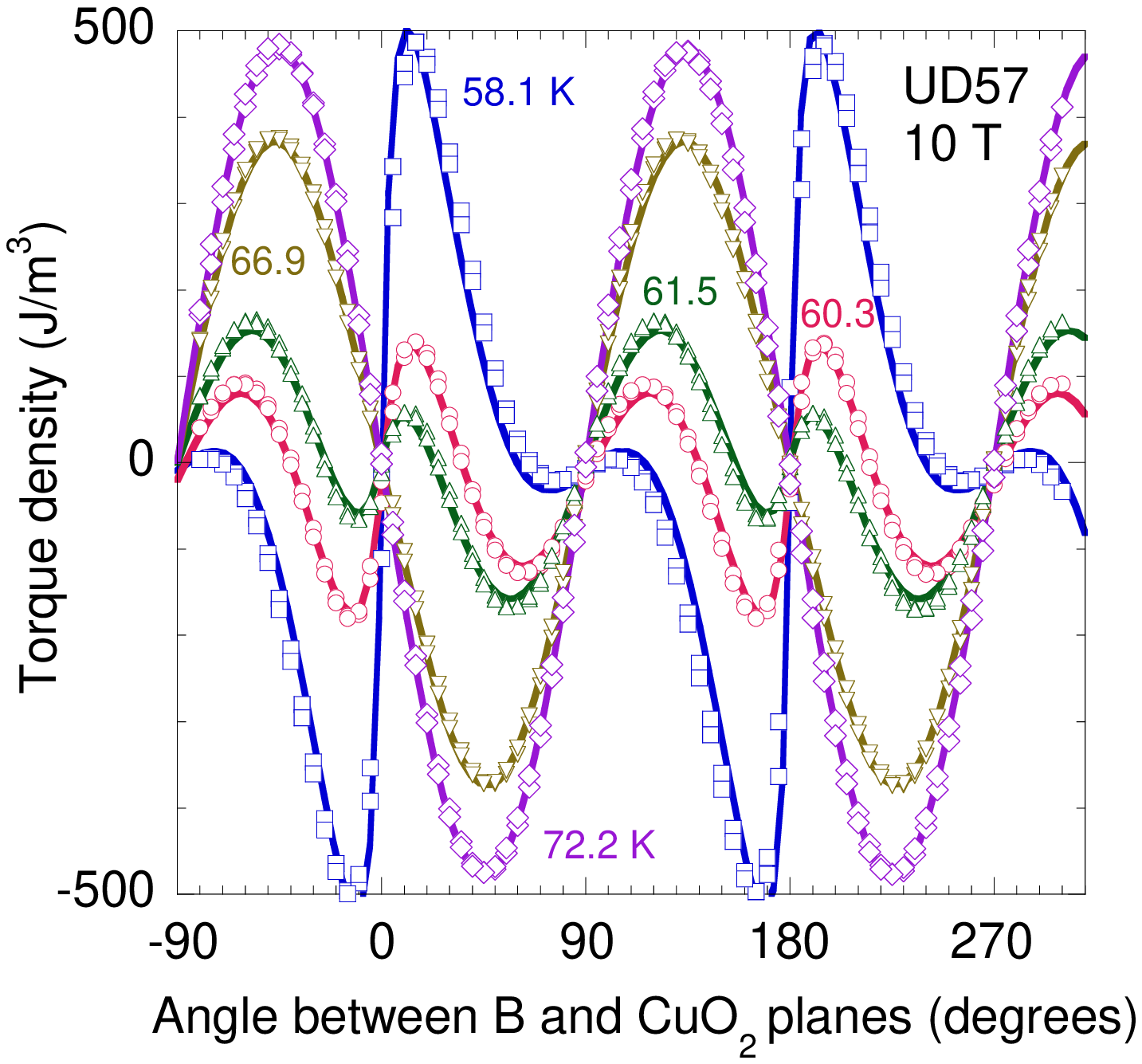}
\caption{ Color online.  Angular dependence of the torque density
 for the UD57
YBa$_{2}$Cu$_{3}$O$_{6.5}$ crystal in 10 T at
 $T$= 58.1, 60.3, 61.5, 66.9  and 72.2~K after  correcting for a
   fixed instrumental offset of 10$^\circ$ and subtracting the gravitational term \cite{exptldetails}. The solid lines
 show single parameter fits to the formula
 for 2D GF derived from Eq.~2 plus $\chi_D^N(T)$ shown in Fig.~2a.  Note the $\sin2\theta$ behavior at higher $T$. }
 \label{rawdata1}
\end{figure}

Fig.~2a shows  $\chi_D(T)$ obtained from $\sin2\theta$ fits for three doping levels at
high enough $T$ so that $M$ remains $\propto B$. The solid lines for OD89  and UD57
are fits up to 300~K that include $\chi_c^{FL}(T)$ from Eq.~1, with the strong cut-off
described below,  plus the normal state background anisotropy $\chi_D^N(T)$ which
arises from the $g$-factor anisotropy of the Pauli paramagnetism \cite{KokanovicEPL}.
For UD crystals the $T$-dependence of $\chi_D^N(T)$ is caused by the pseudogap, see
footnote~\onlinecite{CDWnote}, plus a smaller contribution from the electron pocket
\cite{KokanovicEPL} observed in high field quantum oscillation studies
\cite{TailleferReview}. We used the same pseudogap energies ($k_BT^*$) and  other
parameters defining $\chi_D^N(T)$ as in our recent work on larger single crystals
\cite{KokanovicEPL}, e.g. $T^*$ = 435 K for UD57. OD89 has no pseudogap and presumably
no pockets, so we represent the weak variation of $\chi_D^{N}(T)$ with $T$ by the
second order polynomial shown in Fig.~2a.

 \begin{figure}
\includegraphics[width=7.5cm,keepaspectratio=true]{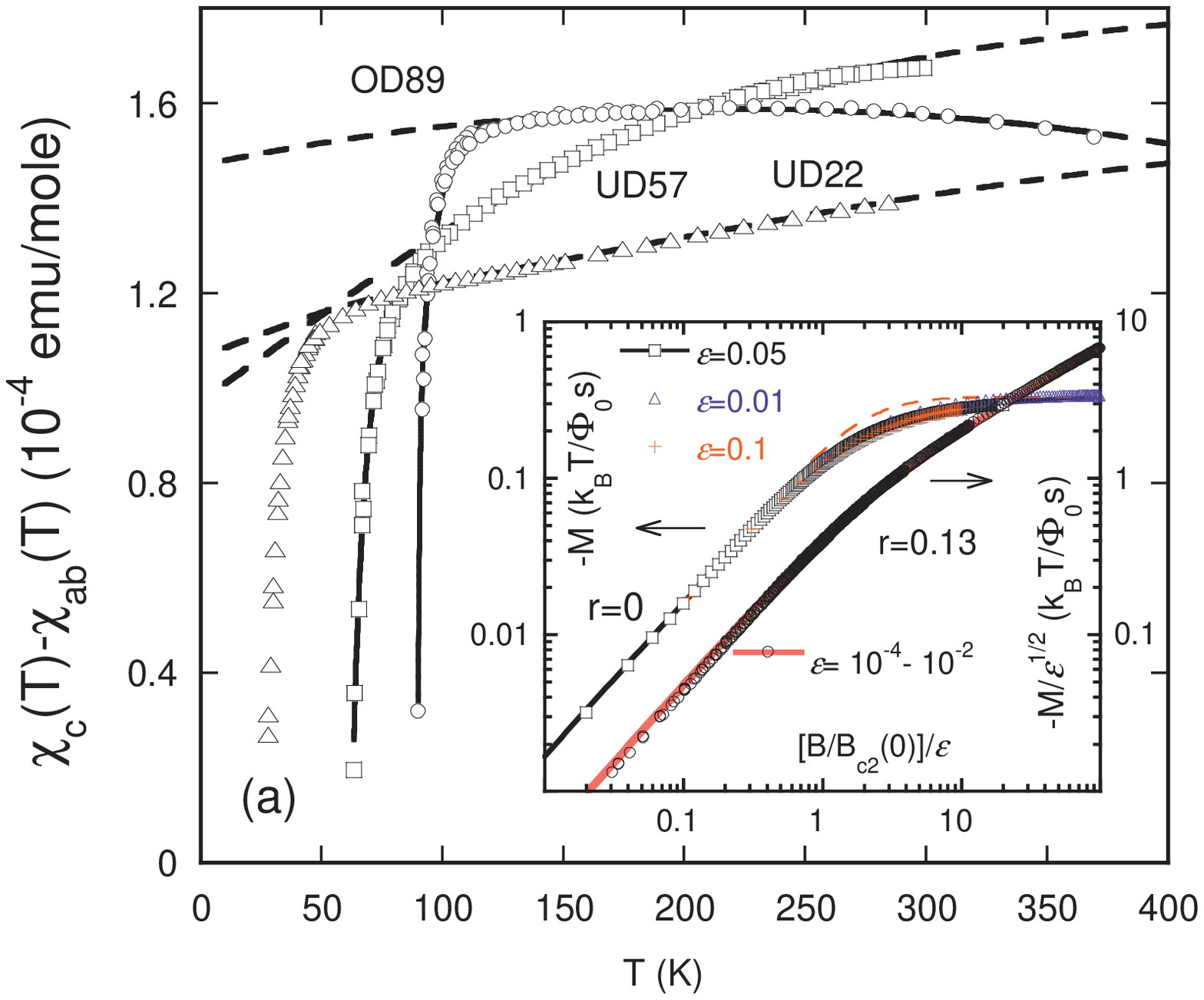}
\includegraphics[width=7.0cm,keepaspectratio=true]{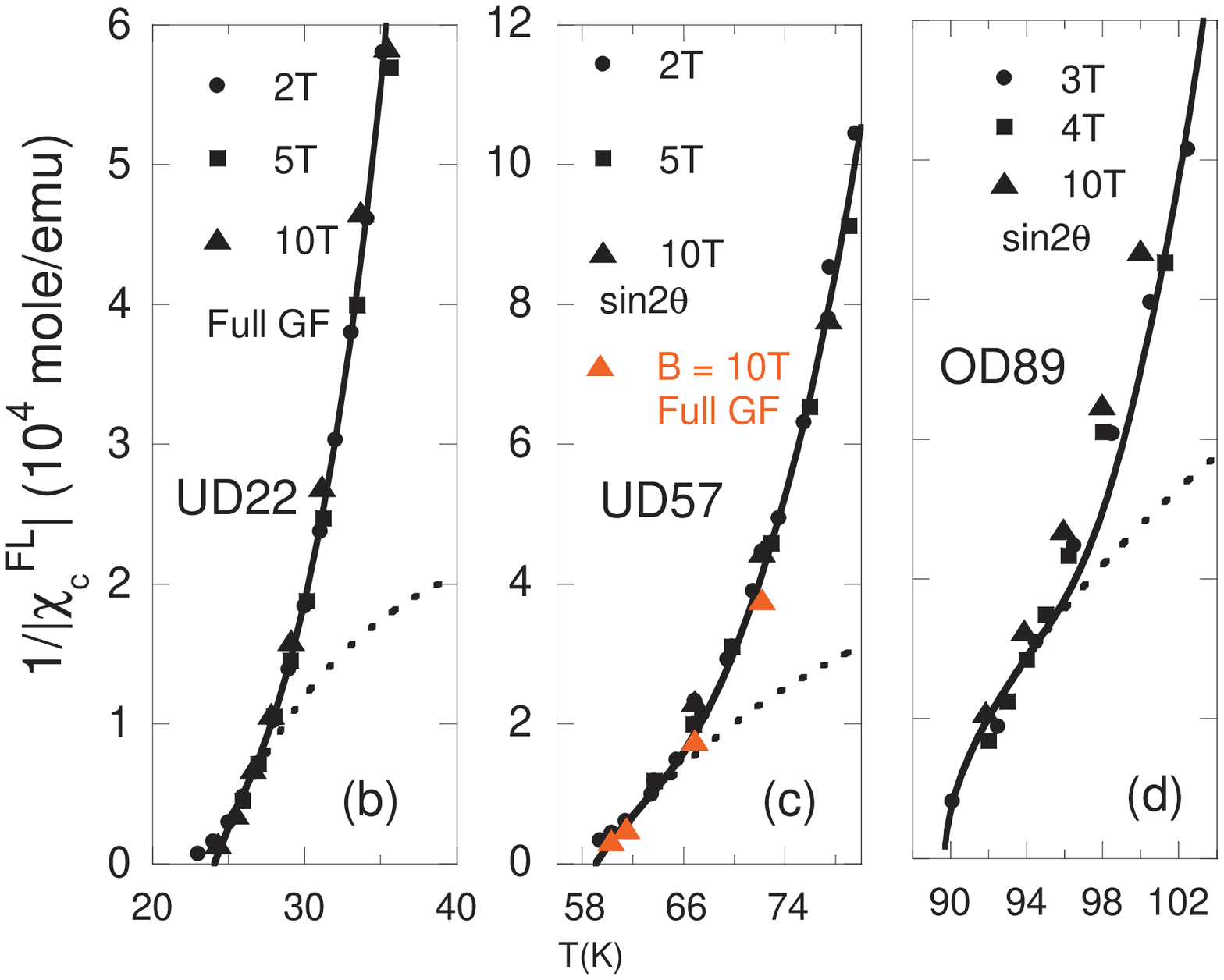}
\caption{Color online: (a) Main: $\chi_D(T)$ for the three crystals, solid lines show
fits to $\chi_c^{FL}(T)+\chi_D^N(T)$ for OD89 and UD57, dashed lines show
$\chi_D^N(T)$. Insert: Symbols show $M$ calculated for various values of $\epsilon$,
using Eq.~2, when the anisotropy parameter $r\equiv(2\xi_c(0)/s)^2$ = 0. For $r$= 0.13
symbols show $M$ given by  the 2D-3D form of Eq.~2, which contains $r$ and an extra
integral \cite{Larkin}. The lines show formulae used \cite{insertnote} to represent
these values of $M$ when fitting $\tau(\theta)$ data.\\(b) to (d) - plots of
  $1/|\chi_c^{FL}(T)|$  vs. $T$ for the three
crystals. GF fits based on Eq.~1, are shown by short dashed lines, without a cut-off
and by solid lines, with a strong cut-off \cite{cutoffnote}. Red  triangles for UD57
show $\xi_{ab}(0)^2/\epsilon$  obtained by fitting $\tau(\theta)$ to the full 2D GF
formula when $M(B)$ is non-linear, and converted to
 $1/|\chi_c^{FL}(T)|$ using Eq.~1. For UD22 the full GF formula was used for all the points shown in Fig. 2b.}
 \label{chidata}
\end{figure}

 Figs.~2b to 2d show plots of $1/|\chi_c^{FL}(T)|$ vs. $T$ where
$\chi_c^{FL}(T) \equiv \chi_D(T)- \chi_D^{N}(T)$.  The short-dashed lines for UD22 and
UD57 in Figs.~2b and 2c show the contribution from Eq.~1 in the 2D limit ($\gamma
\rightarrow\infty$)  with the two adjustable parameters $T_c^{MF_1}$ and $\xi_{ab}(0)$
given in Table 1. The solid lines show the effect of the same type of cut-off used in
previous studies of the the conductivity $\sigma_{ab}^{FL}(T,B)$, as summarized in
footnote~\onlinecite{cutoffnote}. For OD89 we use the full 2D-3D form of Eq.~1 with
$\xi_{ab}(0)$ = 1.06 nm and $\gamma$ = 5, \cite{Babic} shown by the short-dashed line,
with the solid line again including the cut-off \cite{cutoffnote}. The high quality of
these fits could be somewhat fortuitous in view of our neglect of any charge density
wave (CDW) \cite{CDWnote}, but  other subtraction procedures give similar values of
$1/|\chi_c^{FL}(T)|$.  Heat capacity studies give a very similar value $\xi_{ab}(0)$ =
1.12 nm for OD88 YBCO \cite{LoramPhilMag} while our values for UD57 and UD22 agree
with previous work \cite{Alloul,AndoHc2} for the same $T_c$ values. For UD57, setting
$\gamma = 45$ \cite{Pereg}, rather than the 2D limit of Eq.~1 ($
\gamma\rightarrow\infty$) has no significant effect.

As the critical region is approached from above $T_c$ the exponent of $\xi_{ab}(T)$ is
expected to change from the MF value of -1/2 to the 3D-XY value of -2/3
\cite{Bulaevskii}. It is very likely that this will also apply to strongly 2D
materials, including UD57, since heat capacity data above and below $T_c$
\cite{LoramLnt} do show the $\ln|\epsilon|$ terms associated with the 3D-XY model. We
have addressed this by repeating our GF fits in Figs.~2b and 2c with
$\epsilon\geq0.20$ (UD22) or $0.15$ (UD57) without altering the cut-off
\cite{cutoffnote}. The only significant change is that $\xi_{ab}(0)$ becomes 15$\%$
larger for UD57.  For OD89 fits with $T_c^{MF_1}$ = 90~K and $\epsilon\geq0.05$ do not
alter $\xi_{ab}(0)$ within the quoted error.  This is expected since the  width of the
critical region for OD89 is much smaller than for OP YBCO \cite{Kamal,Meingast}
because of the extra 3D coupling  from the highly conducting CuO chains
\cite{LoramPhilMag}.
 \begin{figure}
\includegraphics[width=7.0cm,keepaspectratio=true]{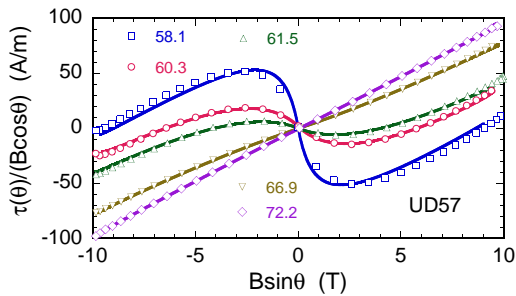}
\caption{ Color online: Magnetic field dependence of the magnetization obtained from
the
 torque data for
UD57 at $T$ = 58.1, 60.3, 61.5, 66.9 and 72.2 K. Solid lines show fits to the 2D  GF
formula for $M$ plus the same normal state contribution used in Figs.~1,~2a and~2c.}
 \label{magdata}
\end{figure}

Fig.~3  shows plots of $\tau/B\cos\theta$ vs $B\sin\theta$ at fixed $T$ for UD57. We
use this representation of the data and MKS units, A/m, for  comparison with
Ref.~\onlinecite{LuLi}. If $\chi_D^{N}(T)$ is  subtracted, which has not been  done
for Fig.~3, then  since  $M_{ab}^{FL}$  is small this would be the same as plotting
$M_c^{FL}$ vs. $B\parallel c$. Near $T_c$ there is clear non-linearity  which is
remarkably consistent with  GF in the 2D limit, for which the free energy density at
all $B$ is~\cite{Larkin}:
\begin{equation}
{F = \frac{k_BT}{2\pi\xi_{ab}^2s}\{b\ln[\Gamma(\frac{1}{2}+
\frac{\epsilon}{2b})/\sqrt{2\pi}]+ \frac{\epsilon}{2}\ln(b)\} } \label{2DFree}
\end{equation}
using the standard  $\Gamma$ function, with $b= B/\tilde{B}_{c2}(0)$, where
 $\tilde{B}_{c2}(0)=\Phi_0/2\pi\xi_{ab}(0)^2$, and as before
$\epsilon=\ln(T/T_c^{MF_1}(B=0))$. The magnetization $M=-\partial F/\partial B$
obtained by numerical differentiation of Eq.~2 for three typical values of $\epsilon$
is shown in the insert to Fig.~2a.  $M$ scales with $b/\epsilon$ to within a few $\%$
and for $0.01<\epsilon<1$ can be adequately represented by the simple formula
$-bk_BT/[\Phi_0s(3b+6\epsilon)]$, that has  a single unknown parameter
$\xi_{ab}(0)^2/\epsilon$. We note that  GF formulae will be approximately valid  in
the crossover region to 3D-XY behavior \cite{Bulaevskii}, because to first order the
main effect is the change in the exponent of $\xi_{ab}(T)$.

Figs.~1 and~3 show that this formula fits  our data for UD57 very well and
importantly,
 as shown by the red triangles in Fig.~2c, the corresponding values of  $1/\chi_c^{FL}(T)$ obtained
 via Eq.~1 agree well with points from $\sin2\theta$ fits at lower $B$ or higher $T$. For
OD89 strong deviations from $\sin2\theta$ behavior only occur within $\sim$ 1 K of
$T_c$ and these \cite{OD89note} are not properly described by GF theory.  For  UD22
there were
 small jumps in $\tau(\theta)$ at $\theta =0$ between 35 and 26 K of size $M_c =
0.01-0.03k_BT/(3\Phi_0s)$ that were fitted by including an extra contribution from
Eq.~2 in the $\epsilon\ll b$ limit.  This is  ascribed to small regions, 1 to 3$\%$ of
the total volume, with higher $T_c$ \cite{Lascialfari} that are   not detected in
low-field measurements of $T_c$ because they are much smaller than the London
penetration depth.  Fig.~2b  shows that  the values of $\xi_{ab}(0)^2/\epsilon$ [or
equivalently $1/\chi_c^{FL}(T)$] obtained from full GF fits to $\tau(\theta)$ data at
2, 5 and 10~T agree well, which supports this conclusion.

The good description of our data by this GF analysis suggests that the high critical
fields proposed in Refs.~\onlinecite{LuLi,Xu,Wang06} for $0.01<\epsilon\lesssim 0.2$
are \underline{not} associated with vortex-like excitations. In the present picture 2D
GF give $M_c^{FL} \simeq -0.33 k_BT/\Phi_0s = -0.112$~emu/cm$^3$ or -112 A/m at 60~K
for $B\gtrsim \phi_0/[2\pi\xi_{ab}(T)^2]$. We expect this to  be suppressed for
$B\gtrsim B_{c2}(0)$ where the magnetic length becomes smaller than $\xi_{ab}(0)$ and
the slow spatial variation approximation of GL theory breaks down. However it may also
fall when $\epsilon\gtrsim0.1$ because of the GF cut-off discussed below. So in the
first approximation the high fields are $\simeq
 B_{c2}(0)$.   Precise analysis of these
effects at very high fields might need to allow for small changes in $\chi_D^N(T)$
with $B$ that depend on the ratio of the Zeeman energy to  the pseudogap. We note that
the present results are consistent with a recent study of $B_{c2}$ for YBCO
\cite{Taillefer2} and that recent torque magnetometry data \cite{Barisic2012} for
HgBa$_2$CuO$_{4+x}$ and other single layer cuprates,  show similar exponential
attenuation factors to those for YBCO \cite{Alloul,cutoffnote}.

An intriguing question about the present results and those of Ref.~\onlinecite{Alloul}
is the origin of the strong cut-off in the GF above $\sim1.1T_c$. If  the weakly
$T$-dependent $\chi^N_D(T)$ behavior for OD89 shown in Fig.~2a  is correct then our
$\chi^{FL}_D(T)$ data and  $\sigma_{ab}^{FL}(T)$ \cite{Alloul} both decay as
$\exp[-(T-1.08T_c)/T_0]$ above $T\sim1.08T_c$ with $T_0\sim9$~K. If instead
$\chi^N_D(T)$ were  constant below 200~K then our $\chi^{FL}_D(T)$ data would give
$T_0\sim$25~K, a slower decay than Ref.~\onlinecite{Alloul}. In either case the
presence of this cut-off for OD YBCO   rules out explanations connected with the mean
distance between carriers. This is much less than $\xi_{ab}(0)$ for  hole
concentrations of $\simeq1.2$ per CuO$_2$ unit, the value found directly from
 quantum oscillation studies of OD
Tl$_2$Ba$_2$CuO$_{6+x}$ crystals \cite{Rourke}.

Assuming there are no unsuspected effects caused by $d$-wave pairing, one hypothesis
is that the GF and possibly $T_c$ itself are suppressed by inelastic scattering
processes. In a quasi-2D Fermi liquid the inelastic mean free path, $l_{in}$, can be
found from the $T$-dependence of the electrical resistivity and the circumference of
the Fermi surface.  For OD YBCO the measured $a$-axis resistivity \cite{AndoHc2} gives
$l_{in}$ = $2.5(100/T)$ nm, but values for UD samples are less certain because of the
pseudogap. The
 BCS relation  $\xi_{ab}(0)= \hbar v_F/\pi\Delta(0)$, where $\Delta(0)$ is the superconducting
  energy gap at $T=0$,  implies
that irrespective of the value of the Fermi velocity $v_F$, the usual pair-breaking
condition  for significant inelastic scattering, $\hbar/\tau_{in} \gtrsim \Delta(0)$
is equivalent to $l_{in}\lesssim\pi\xi_{ab}(0)$.  Taking $\xi_{ab}(0)$ from Table I
and the above value of $l_{in}$ shows that this is satisfied at 100 K for OD YBCO. So
some suppression of GF and indeed $T_c$ by inelastic scattering is entirely plausible.
If $T_c$ is suppressed then $\Delta(T)$ will fall more quickly than BCS theory as
$T_c$ is approached from below, which would affect the analysis of
Ref.~\onlinecite{Tallon2011}.

 Another possibility \cite{Alloul} which
might account for the observations,
  is that the pairing strength
itself falls sharply outside the GL region, for example when the in-plane coherence
length becomes comparable to, or less than, the correlation length of spin
fluctuations. From Figs.~2b to~2d we can read off  the values of $T$  where the solid
and dashed lines differ by (say) a factor of two. At these points
$\xi_{ab}(T)\equiv\xi_{ab}(0)/\ln(T/T_c^{MF_1})$ = 15.6, 9.5 and 7.9 nm for UD22, UD57
and OD89 respectively. Neutron scattering studies \cite{Hayden,Stock} typically give a
full width half maximum of 0.17$\frac{2\pi}{a}$ for the scattering intensity from spin
fluctuations. Although this does vary with composition and scattering energy it
corresponds to a correlation length \cite{Kittel} of just over 6 lattice constants,
$a$, or 2.5 nm, similar to  $\xi_{ab}(0)$ but much smaller than the $\xi_{ab}(T)$
values for which $\chi_c^{FL}$ is reduced by a factor two. It remains to be seen
whether theory could account for this.

In these two pictures the effective $T_c$ describing the strength of the GF would fall
for $T>1.1T_c$ either because of inelastic scattering or because of a weakening of the
pairing interaction. If it could be shown theoretically that $\tilde{B}_{c2}(0)$ falls
in a similar way, this would account naturally for the fact \cite{Alloul} that the
magnetic  fields needed to destroy the GF fall to zero in the temperature range
120-140 K, where the fluctuations become very small.  In summary, Gaussian
superconducting fluctuations, plus a strong cut-off that seems to be linked to a
reduction in the effective value of $T_c$, provide a good description of the
diamagnetism of our superconducting cuprate crystals above $T_c$.

\begin{table}
\begin{ruledtabular}
\begin{tabular}{l|c|c|c|c|c}
$Sample$&$^\S T_c $&$T_c^{MF_1}$&$\xi_{ab}(0)$&$0.59\tilde{B}_{c2}(0)^{\ddag}$&$\Delta(0)^{\dag}$\\
$ $&$ (K)$&$  (K)$&$(nm)$&$ (T)$&$ (K)$\\
\hline OD89 &$89.4$&$89.7$ &$1.06\pm0.1$&$173$&$448$\\
\hline UD57 &$56.5$&$59$ &$2.02\pm0.1$&$48$&$234$\\
\hline UD22 &$21.6$&$24 $&$4.5\pm0.5$&$10$&$105$\\
\end{tabular}
\end{ruledtabular}
\caption{Summary of results.  $^\S T_c$ defined by sharp onsets of SQUID signal at 10G
and torque data at $\pm$50G.  $^{\ddag}$2D clean limit formula \cite{Larkin} for
$B_{c2}(0)$. $^{\dag}$From the BCS relation $\xi_{ab}(0)=\frac{\hbar
v_F}{\pi\Delta(0)}$, which may not hold exactly for $d$-wave pairing, with
$v_F$=2x10$^{7}$ cm/sec. } \label{summary}
\end{table}

We are grateful to D. A. Bonn, A. Carrington, W. N. Hardy, G. G. Lonzarich, J. W.
Loram and L. Taillefer for several helpful comments. This work was supported by EPSRC
(UK), grant number EP/C511778/1 and the Croatian Research Council, MZOS project
No.119-1191458-1008.


\begin{thebibliography}{99}
\bibitem{exptldetails}The crystal is glued to the end of a commercial piezolever
with its CuO$_{2}$ planes parallel to the flat surface of the lever. A dummy lever compensates background
magneto-resistance signals, using a 3-lead Wheatstone bridge circuit driven by a
floating 77 Hz current source. The chip is mounted on a single-axis rotation stage
inside a He$^4$ cryo-magnetic system providing stable temperatures from 1.4 K up to
400 K and fields up to 15 T. The bridge signal arising from the gravitational torque
on the crystal when the sample stage is rotated in zero magnetic field gives the
$T$-dependent sensitivity of the piezolever. Because the masses of the glue and the
lever are much less than that of the crystal, the calibration constant relating the
out-of balance bridge signal to the angular dependent torque density $\tau(\theta)$ in
J/m$^3$ or
 $\chi_D(T)$ \cite{emumole}, only depends on
the distance between the center of mass of the crystal and the base of the lever at
the silicon chip, measured to $\pm5\%$.
\bibitem{cutoffnote} We fitted the normalized
 $\sigma_{ab}^{FL}(T)$ data  in Fig.~25 of Ref.~\onlinecite{Alloul}
to an empirical formula $(\exp[(T- \alpha T_c)/\beta] +1)^{-0.1}$  which is $\approx
1$ for $\epsilon\lesssim0.1$ and $\approx\exp[-(T-\alpha T_c)/10\beta]$ at higher $T$.
This formula was used to cut off $\chi_c^{FL}(T)$ with $\alpha$ =  1.078, 1.1 and 1.12
and $\beta$ = 0.869, 1.234 and 0.70~K for OD89, UD57 and UD22 respectively and
$T_c=T_c^{MF_1}$ shown  in Table 1.  For OD89, $\alpha$ and $\beta$ values correspond
 to  OD92.5 in Ref.~\onlinecite{Alloul},  for UD57 we used
 UD85 data in Ref.~\onlinecite{Alloul} which are similar to UD57 but have less scatter.
\bibitem{insertnote}
 The solid line for $r=0$ shows our empirical
 2D formula $b/(3b+6\epsilon)$, where $b=2\pi\xi_{ab}(0)^2B/\Phi_0$.
 The dashed line shows the 2D limit of Eq.~1 with $\xi_{eff}(b)$ given
  by $\xi_{eff}(b)^{-4}=\xi_{ab}(T)^{-4}+l_B^{-4}$, where $l_B=(\hbar/eB)^{1/2}$, the formula used to analyze
  Nernst data for NbSi films \cite{Pourret}. For $r=0.13$, $b<r$ and $\epsilon<r$,
our empirical 3D formula is
 $-M/\sqrt{\epsilon}=(k_BT/s\Phi_0)0.68b/\sqrt{\epsilon(b+1.94\epsilon)}$.
\bibitem{OD89note}  Although the 2D-3D form of Eq.~2 \cite{Larkin} with $r=0.13$
describes
 the non-$\sin2\theta$ shape of $\tau(\theta)$
the calculated values of $M\parallel c$ are a factor of 3 too small, and $\epsilon$ is
far too small compared with the low-field transition width arising from inhomogeneity
or strain. This non-GF behavior is ascribed to $T$ being too close to $T_c$.
\bibitem{emumole} Units: 1 J/m$^3$ = 10 ergs/cm$^3$ and using CGS units for
 $\tau(\theta)= \frac{1}{2}\chi_DB^2\sin2\theta$ with  $B$ in gauss gives $\chi_D$ in
 emu/cm$^3$. Complete flux exclusion corresponds to $\chi$ = -1/4$\pi$ emu/cm$^3$, or $\chi$ = -1 in  MKS units.  For YBCO $\chi_D$ in
 emu/cm$^3$, is multiplied by the volume per mole, 666/6.38 cm$^3$ to convert to emu/mole.
\end{thebibliography}
\end{document}